\documentclass[fleqn,usenatbib]{mnras}

\usepackage{newtxtext,newtxmath}

\usepackage[T1]{fontenc}
\usepackage{ae,aecompl}

\usepackage{subcaption}

\usepackage{xcolor}
\usepackage{graphicx}	
\usepackage{amsmath}	

\usepackage{epsfig}
\usepackage{rotating}
\usepackage{color}
\usepackage{tabularx}
\usepackage{multirow}
\usepackage{graphics}
\usepackage{epsfig}
\usepackage{epstopdf}
\usepackage[para,online,flushleft]{threeparttable}
\usepackage[ampersand]{easylist}
\definecolor{navy}{rgb}{0,0.3,0.7}
\definecolor{forrest}{rgb}{0,0.55,0.25}
\definecolor{tyrianpurple}{rgb}{0.4, 0.01, 0.24}
\definecolor{coralred}{rgb}{1.0, 0.25, 0.25}
\usepackage{hyperref}
\hypersetup{
    colorlinks=true,
    linkcolor=blue,
    filecolor=blue,      
    urlcolor=blue,
    citecolor=blue,
    filecolor=blue,
}
\title[Optical and HI emission from G108.2$-$0.6]{Discovery of optical emission from the supernova remnant G108.2$-$0.6 and its atomic environment}

\author[Payl{\i} et al.]{{\color {black} G.~Payl{\i}$^{1}$\thanks{E-mail:{\color  {blue}gunaypayli@gmail.com} (GP)},
H.~Bak{\i}\c{s}$^{1}$\thanks{{\color {blue}corresponding author: hicranbakis@akdeniz.edu.tr} (HB)}, 
E.~Aktekin$^{2}$,
H.~Sano$^{3}$, and
A.~Sezer$^{4}$}\\
$^{1}$Department of Space Sciences and Technologies, Akdeniz University, 07058, Antalya, Turkey\\
$^{2}$Department of Physics, S\"{u}leyman Demirel University, 32000, Isparta, Turkey \\
$^{3}$Faculty of Engineering, Gifu University, 1-1 Yanagido, Gifu 501-1193, Japan\\
$^{4}$Department of Computer Engineering, Avrasya University, 61250, Trabzon, Turkey\\
}

\date{Accepted XXX. Received YYY; in original form ZZZ}

\pubyear{2023}

\begin{document}
\label{firstpage}
\pagerange{\pageref{firstpage}--\pageref{lastpage}}
\maketitle

\begin{abstract}
We report the first detection of optical emission from the shell-type Galactic supernova remnant (SNR) G108.2$-$0.6. We obtained H$\alpha$ images and long-slit spectra using the 1.5-m RTT150 telescope to examine the morphological and spectral characteristics of the SNR. We detected several filaments along its north and south regions, which is consistent with its SNR nature. The spectra exhibit [S\,{\sc ii}]/H$\alpha$ ratios in the range of 0.4$-$1.1, indicating emission from shock-heated gas. The oxygen doublet emission lines [O\,{\sc i}]$\lambda$6300, $\lambda$6363 detected in the south region also support the indicator of the presence of shocks. We estimate the electron density using the [S\,{\sc ii}] 6716/6731 ratio ranging from 15 to 1800 cm$^{-3}$. The spectra show a relatively low shock velocity of $V_{\rm s}$ $\sim$ 80 km s$^{-1}$ with the pre-shock cloud density of $n_{\rm c}$ $\sim$18$-$57 cm$^{-3}$. The H$\alpha$/H$\beta$ ratios show significant variation across the observed regions with extinction $E(B-V)$ ranging from 0.22 to 1.65. We also analyzed the archival H\,{\sc i} data and estimated the kinematic distance to the SNR of $\sim$0.8 kpc and dynamical age as $\sim$70$\pm$10 kyr of G108.2$-$0.6.

\end{abstract}

\begin{keywords}
ISM: individual objects: G108.2$-$0.6 $-$ ISM: supernova remnants $-$ H\,{\sc ii} regions  $-$  atomic data
\end{keywords}



\section{Introduction}
Supernova remnants (SNRs) provide us with important information on the supernova explosion mechanism, properties of the surrounding medium, and their interaction (see \citealt{Vi20}, for a review). Most SNRs have been detected in radio wavelengths due to their nonthermal synchrotron emission (e.g. \citealt{Du15, Gr19}).  Optical observations of SNRs allow us to inspect the SNR properties such as the morphology, the properties of the shocked gas, and the density of the ambient medium (e.g. \citealt{St08, Sa13, Fe20, Bo22, Do23}). The optical emission of SNRs is dominated by line emission, in particular H$\alpha$, and forbidden line emission from oxygen ([O\,{\sc iii}], [O\,{\sc i}]), nitrogen ([N\,{\sc ii}]), and sulfur ([S\,{\sc ii}]). The main criterion for optical SNR identification is a  line ratio of [S\,{\sc ii}]/H$\alpha$ $\geq$ 0.4, which is useful in separating the shock emission of SNRs from photoionized gas (e.g. \citealt{Fe85}).

G108.2$-$0.6 is a new faint and large (70 arcmin $\times$ 54 arcmin) shell-type radio SNR discovered in the Canadian Galactic Plane Survey (CGPS) at 1420 MHz \citep{Ti07}. The SNR has an elliptical shell-type with a spectral index of $\alpha$ = - 0.5 $\pm$ 0.1 ($S \sim \nu^{\alpha}$). The 1420 and 408 MHz observations show many bright and faint sources around the SNR: bright SNR G109.1$-$1.0 (southeast), faint SNR G107.5$-$1.5 (south), the bright H\,{\sc ii} regions Sh2$-$142 (southwest), Sh2-146 (east) and Sh2-148 (southeast), and the molecular cloud Sh2$-$152 (southeast).  \citet{Ti07} also presented the multi-wavelength H\,{\sc i}, CO, infrared, X-ray, and optical investigations of G108.2$-$0.6. They found that H\,{\sc i} emission associated with the SNR in the radial velocity range is from $-$53 to $-$58 km s$^{-1}$,  and the infrared emission is close to the eastern boundary.  They also reported that there is no molecular material toward the SNR, and the maps of X-ray ({\it ROSAT}) and optical (Palomar Digitized Sky Survey) show no significant emission associated with the SNR. 

The distance to G108.2$-$0.6 was estimated to be 3.2$\pm$0.6 kpc based on the H\,{\sc i} observations \citep{Ti07}. Recently, \citet{Zh20} determined the extinction distance of 1.02$\pm$0.01 kpc from stars in the line of sight. In this work, we continue our search for optical emission from Galactic SNRs (see \citealt{Ba23}), previously discovered in the radio wavelengths, using the 1.5-m RTT150 telescope of T\"{U}B\.{I}TAK National Observatory (TUG)\footnote{\url{https://tug.tubitak.gov.tr}} in T\"{u}rkiye. We discovered optical filamentary and diffuse emission from G108.2$-$0.6 through our search. We perform imaging and spectroscopic observations to investigate the optical properties of the SNR and the ambient medium. We also analyze the H\,{\sc i} data to investigate the atomic environment of G108.2$-$0.6. The structure of the paper is as follows. We present observations and data reduction methods in Section \ref{obs}. Analysis and results are given in Section \ref{analysis}. Based on these results, we infer the optical properties of the SNR and the ambient medium in Section \ref{discuss} and give our conclusions in Section \ref{conc}.

\section{Observations and data reduction}
\label{obs}
\subsection{Optical}
We performed H$\alpha$ imaging of G108.2$-$0.6 with the 1.5-m RTT150 telescope. 
The detector is a CCD camera with 2048 $\times$ 2048 pixels, each of 13.5 ${\mu}$m $\times$ 13.5 ${\mu}$m, covering 11.1 $\times$ 11.1 arcmin$^2$ field of view (FoV). The details of imaging observations are given in Table \ref{Table1}. The imaging data were processed by using standard procedures with \texttt{Image Reduction Analysis Facility} (\texttt{IRAF})\footnote{\url{https://iraf-community.github.io/}}, including bias and flat-field corrections.

\begin{table}
 \caption{Imaging observations of G108.2$-$0.6.}
 \begin{tabular}{@{}p{1.2cm}p{2.5cm}p{1.5cm}p{1.7cm}@{}}
 \hline
Region & R.A. ; Dec.  & Exposure & Observation  \\
ID  &  (J2000)  &    (s)      & date \\
\hline\hline
&\multicolumn{2}{c}{N region}\\
\hline\hline
\multirow{2}{*}{N1 } & \multirow{2}{*}{22 51 40 ; +58 54 54 }  &  3$\times$300&  2023 July 21  \\
 &  &  1$\times$300&  2023 Aug 14  \\
  &  & 3$\times$600 &  2023 Sep 13  \\
\hline
N2  &  22 50 22 ; +58 50 12 & 3$\times$300 &  2023 July 20  \\
 &  & 3$\times$600 &  2023 Sep 13  \\
\hline
N3  &  22 52 58 ; +59 00 12  & 3$\times$300 &  2023 Aug 15 \\
 &  & 3$\times$600 &  2023 Sep 13  \\
\hline \hline
&\multicolumn{2}{c}{S region}\\
\hline\hline
\multirow{2}{*}{S1 } & \multirow{2}{*}{22 55 01 ; +58 11 58}   & 1$\times$300&  2023 July 20  \\
&  &   3$\times$300&  2023 Aug 15  \\
\hline
\multirow{2}{*}{S2 } & \multirow{2}{*}{22 53 43 ; +58 12 12} & 1$\times$300 &  2023 July 20  \\
 &  & 3$\times$300 &  2023 Aug 15  \\

 \hline
 \hline
   &Filter       			& Wavelength    	 & FWHM             \\ 
   &          			&    (nm)	             &   (nm)             \\ 
 \hline
&H$\alpha$     		    &     656.3           &   5                         \\
&H$\alpha$-cont.     	    	&     644.6           &   13                   \\
\hline
\label{Table1}
\end{tabular}
\end{table}

Based on the H$\alpha$ imaging, the long-slit spectra of G108.2$-$0.6 were subsequently obtained with the TFOSC (TUG Faint Object Spectrograph and Camera) mounted at the Cassegrain (f/7.7) focus of the RTT150. The grism-15 was used in the spectral range between 3230 and 9120 {\AA} including H$\beta$, [O\,{\sc iii}], H$\alpha$, [N\,{\sc ii}], and [S\,{\sc ii}] emission lines with the spectral resolution of  $R$ $\sim$ 749. The slit width is 2.38 arcsec, with slit length of 11.1 arcmin and the slit was oriented in the east–west direction. Table \ref{Table2} lists the log of the spectroscopic observations. The data were reduced using standard procedures with \texttt {IRAF}, including bias subtraction, flat-field correction, wavelength, and flux calibrations. The spectrum of an Iron–Argon lamp obtained during the observations was used for wavelength calibration. For flux calibration, we used spectra taken on the same night of spectrophotometric standard star BD+28D4211 \citep{Ok90}.

\begin{table}
\centering
 \caption{Spectroscopic observations of G108.2$-$0.6.}
 \begin{tabular}{@{}p{1.2cm}p{2.6cm}p{1.2cm}p{1.8cm}@{}}
 \hline
Slit  &  R.A. ; Dec.  & Exposure & Observation  \\
 ID & J(2000) &  (s)  & date \\
\hline
&\multicolumn{2}{c}{N region}\\
\hline
N1a  &  22 51 27.5 ; +58 52 09  &  1$\times$1800   & 2023 Jul 21    \\
N1b  &   22 51 56.6 ; +58 52 51 &  1$\times$1800   & 2023 Jul 20    \\
N1c   &   22 52 08.3 ; +58 55 08 &  1$\times$1800   & 2023 Aug 15    \\
\hline
N2a  &  22 50 02.1 ; +58 53 13  &  1$\times$1800   & 2023 Jul 21    \\
N2b  &   22 50 24.1 ; +58 49 48 &  1$\times$1800   & 2023 Jul 21    \\
\hline
&\multicolumn{2}{c}{S region}\\
\hline
S1a  &      22 55 14.8; +58 11 11.9 & 1$\times$1800    &   2023 Aug 14   \\
S1b  &      22 55 21.2; +58 11 22 &  1$\times$1800    &   2023 Aug 14   \\
\hline
S2a  &      22 53 47.8; +58 08 50 &   1$\times$1800   &  2023 Aug 14    \\
S2b &      22 53 41.1; +58 11 40 &   1$\times$1800   &  2023 Aug 14    \\
S2c  &      22 53 48.6; +58 08 44 &   1$\times$1800   &  2023 Aug 15    \\
S2d  &      22 53 38.4; +58 11 51 &  1$\times$1800   &  2023 Aug 15    \\
\hline
\hline
\label{Table2}
\end{tabular}
\end{table}

\subsection{H\,{\sc i}}
We analyzed the H\,{\sc i} line and radio continuum at 1420 MHz taken from a part of the Canadian Galactic Plane Survey (CGPS; \citealt{Ta03}), which was obtained at the Dominion Radio Astrophysical Observatory (DRAO). The angular resolution was 58$\arcsec$ $\times$ 80$\arcsec$ for the H\,{\sc i} data and 49$\arcsec$ $\times$ 68$\arcsec$ for the radio continuum data. The typical noise fluctuation was $\sim$3 K for the H\,{\sc i} data and $\sim$0.3 mJy beam$^{-1}$ for the radio continuum data.    
In Fig. \ref{figure1}, we show the 1420 MHz radio continuum image of G108.2$-$0.6  taken from the CGPS \citep{Ta03}.

\section{Analysis and Results}
\label{analysis}
\subsection{Optical analysis and results}
\subsubsection{Images}
\begin{figure*}
\includegraphics[angle=0, width=12cm]{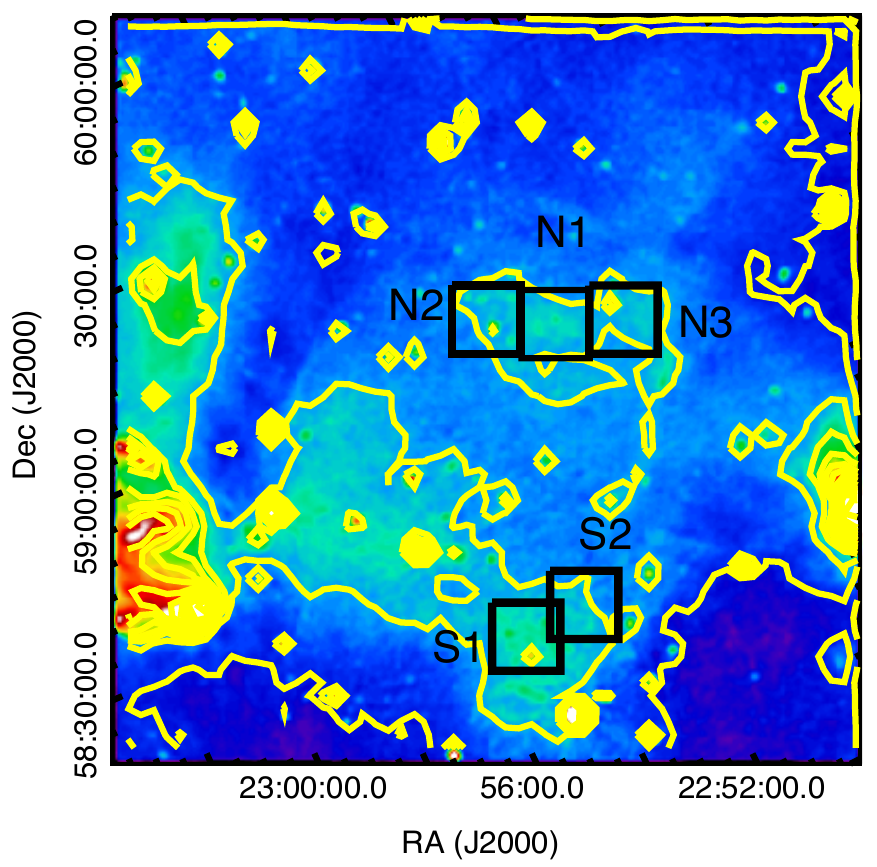}
\caption{The 1420 MHz radio-continuum image of G108.2$-$0.6 taken from the CGPS \citep{Ta03}. The radio contour (yellow) levels range from 2.5 to 30 mJy beam$^{-1}$. The black boxes (11 arcmin $\times$ 11 arcmin) indicate the five regions (NI, N2, N3, S1, and S2) we observed with the RTT150 telescope.}
\label{figure1}
\end{figure*}

\begin{figure*}
\includegraphics[angle=0, width=8.2cm]{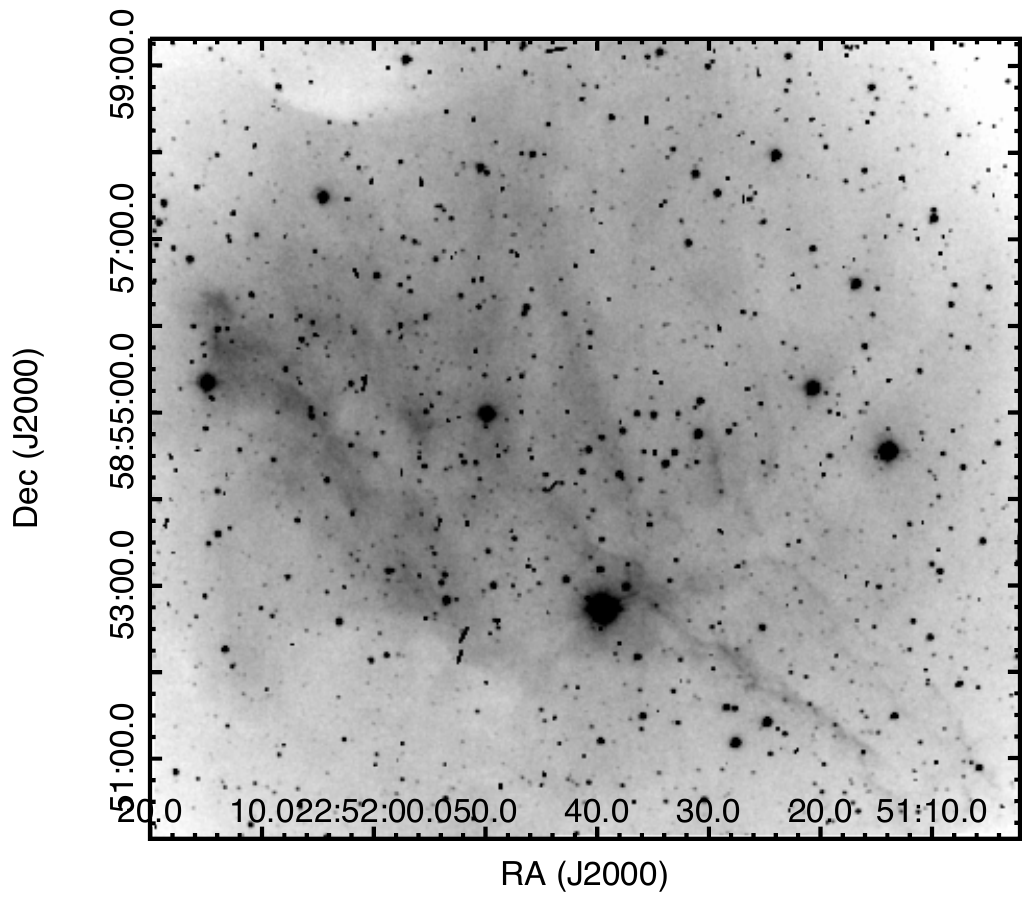}
\includegraphics[angle=0, width=8.2cm]{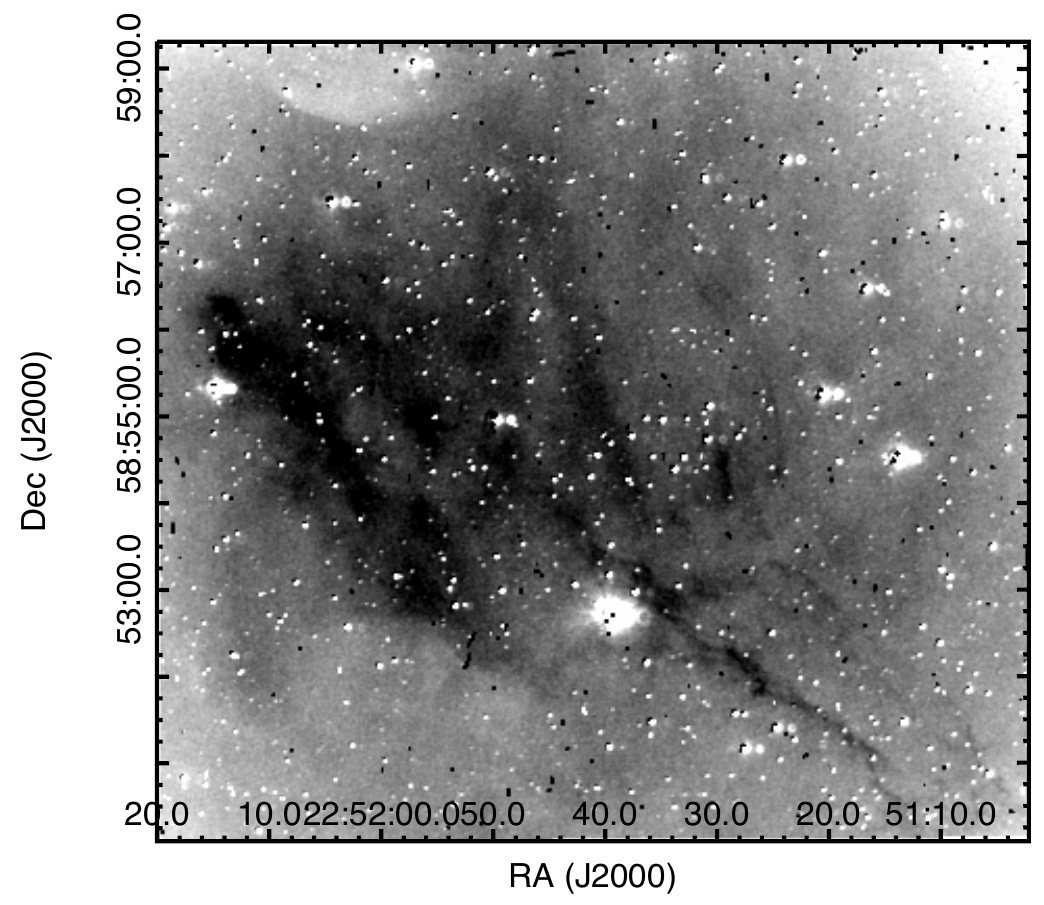}
\includegraphics[angle=0, width=8.2cm]{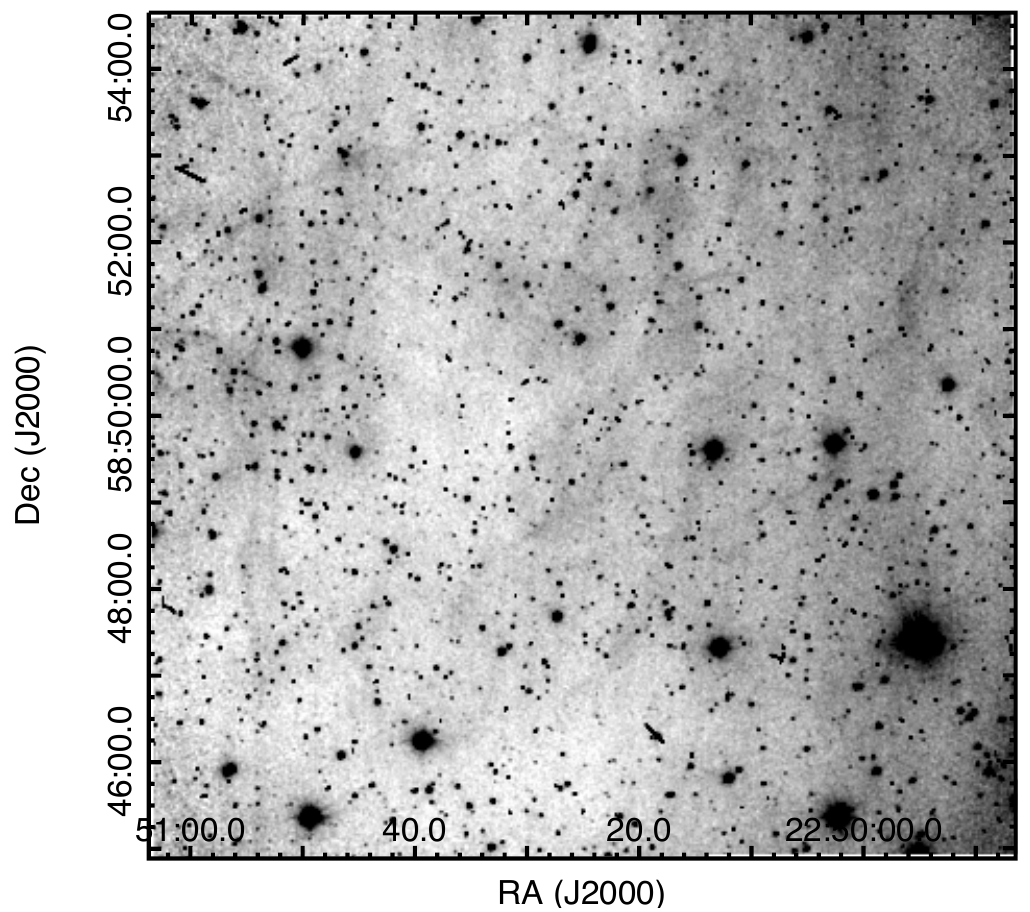}
\includegraphics[angle=0, width=8.2cm]{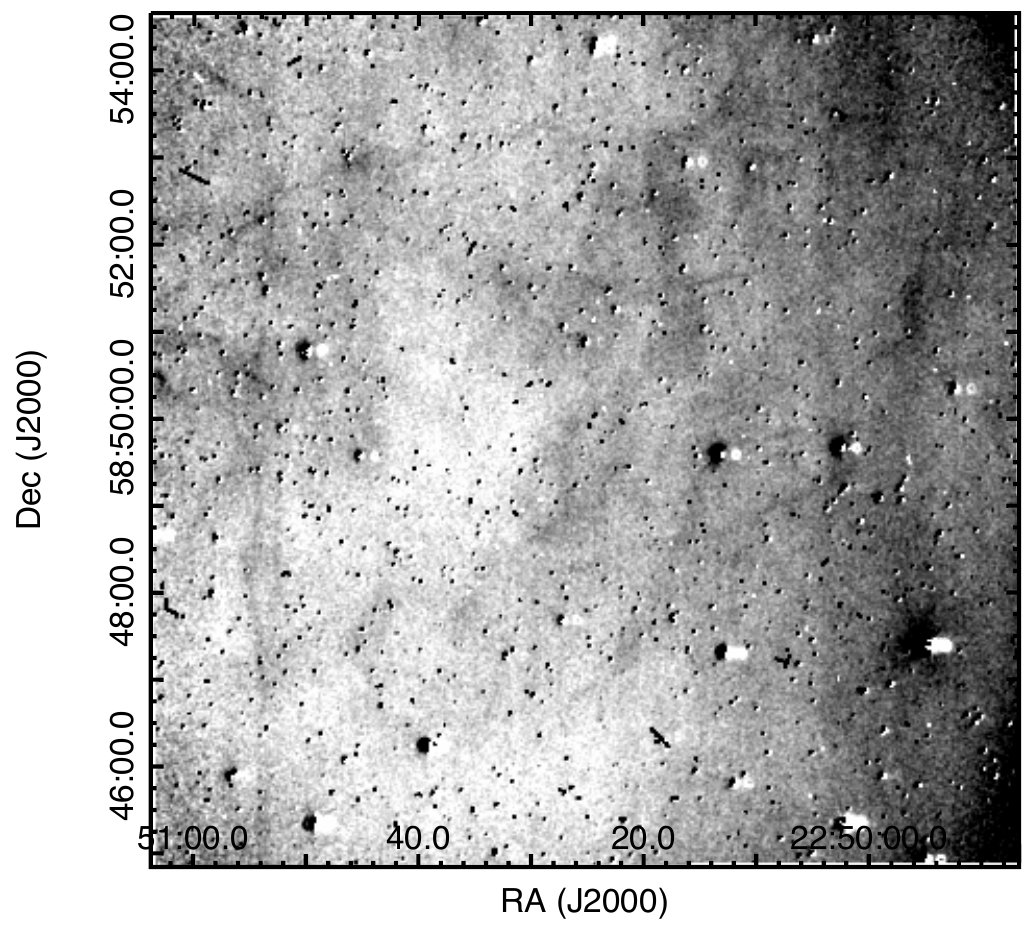}
\includegraphics[angle=0, width=8.2cm]{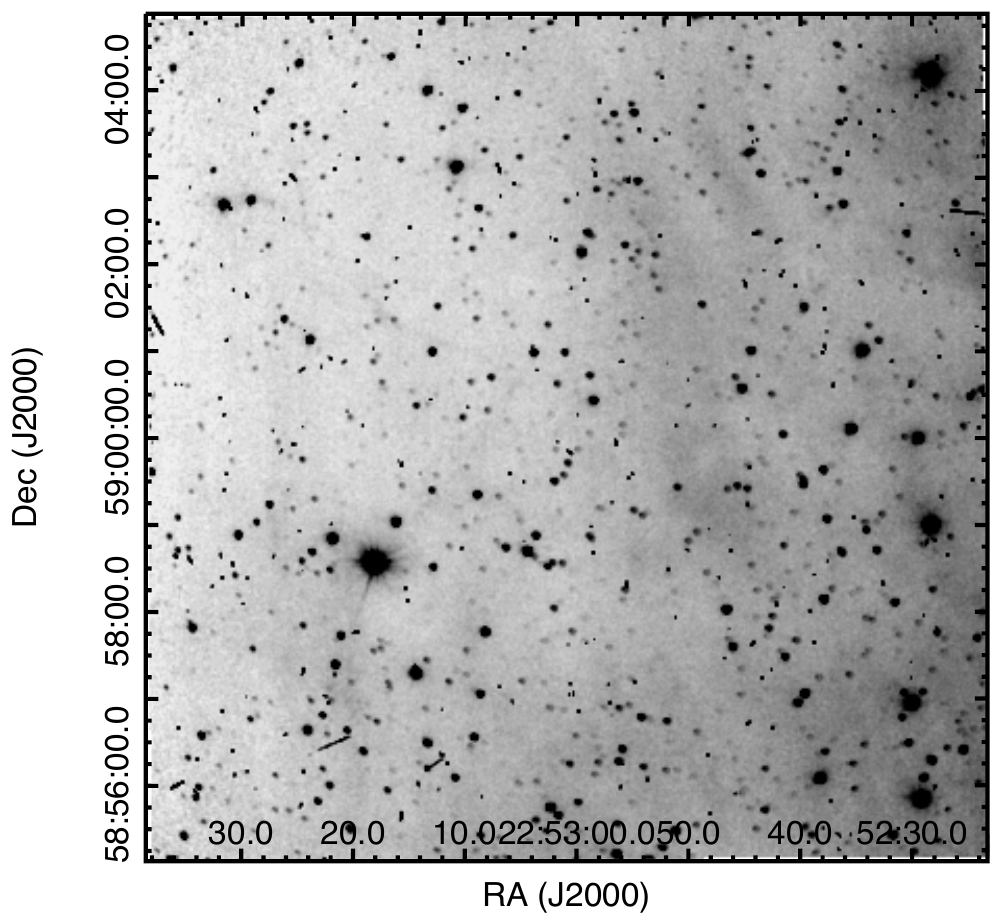}
\includegraphics[angle=0, width=8.2cm]{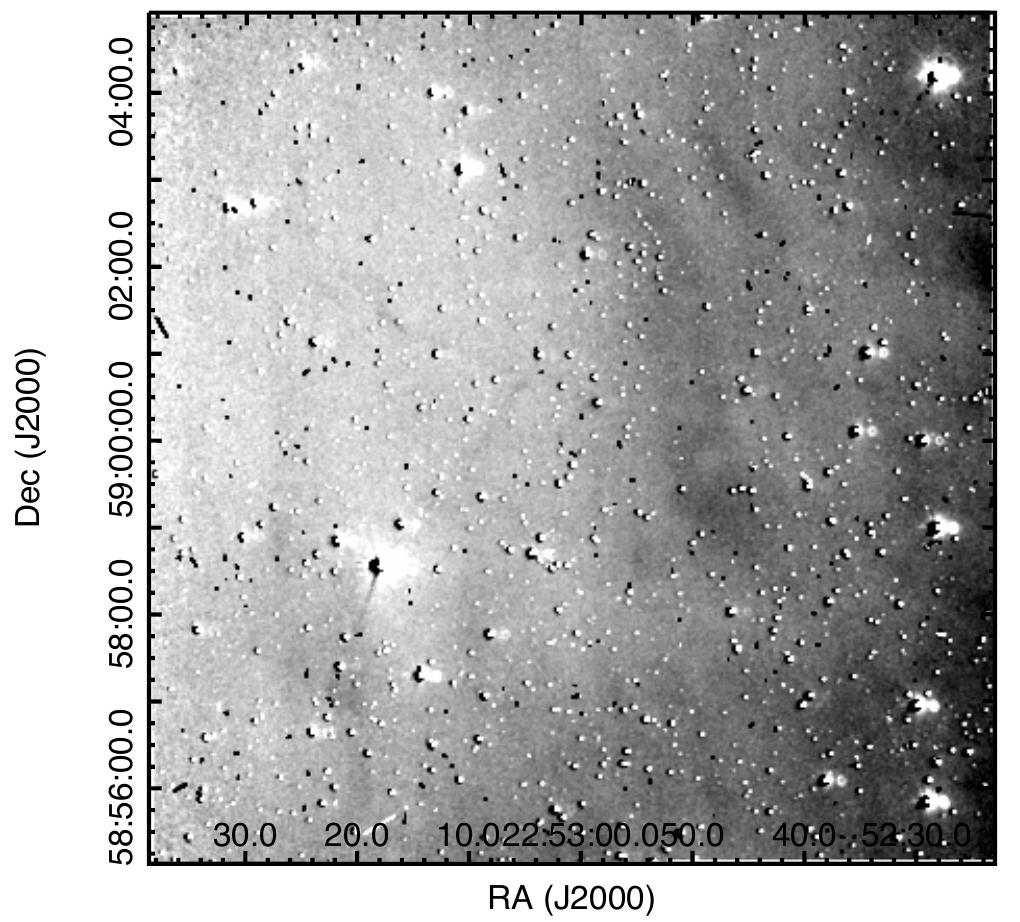}

\caption{Top: The H$\alpha$ (left) and continuum-subtracted H$\alpha$ (right) images of N1 (top panel), N2 (medium panel) and N3 (bottom panel) regions of G108.2$-$0.6.}
\label{figure2}
\end{figure*}

\begin{figure*}
\includegraphics[angle=0, width=8.5cm]{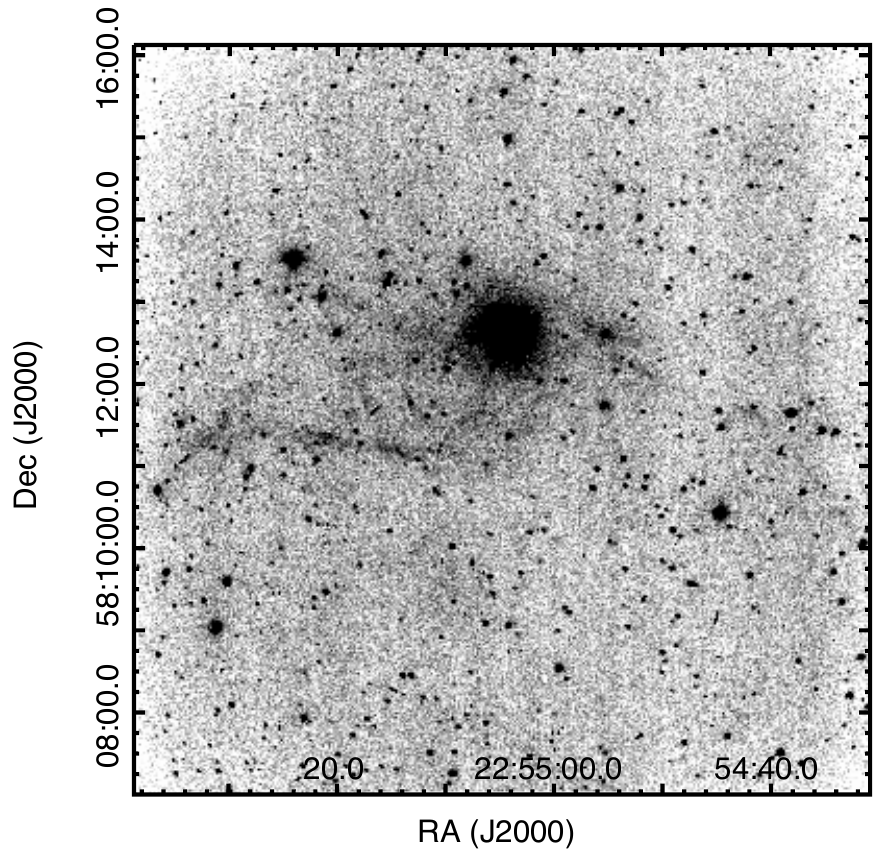}
\includegraphics[angle=0, width=8.5cm]{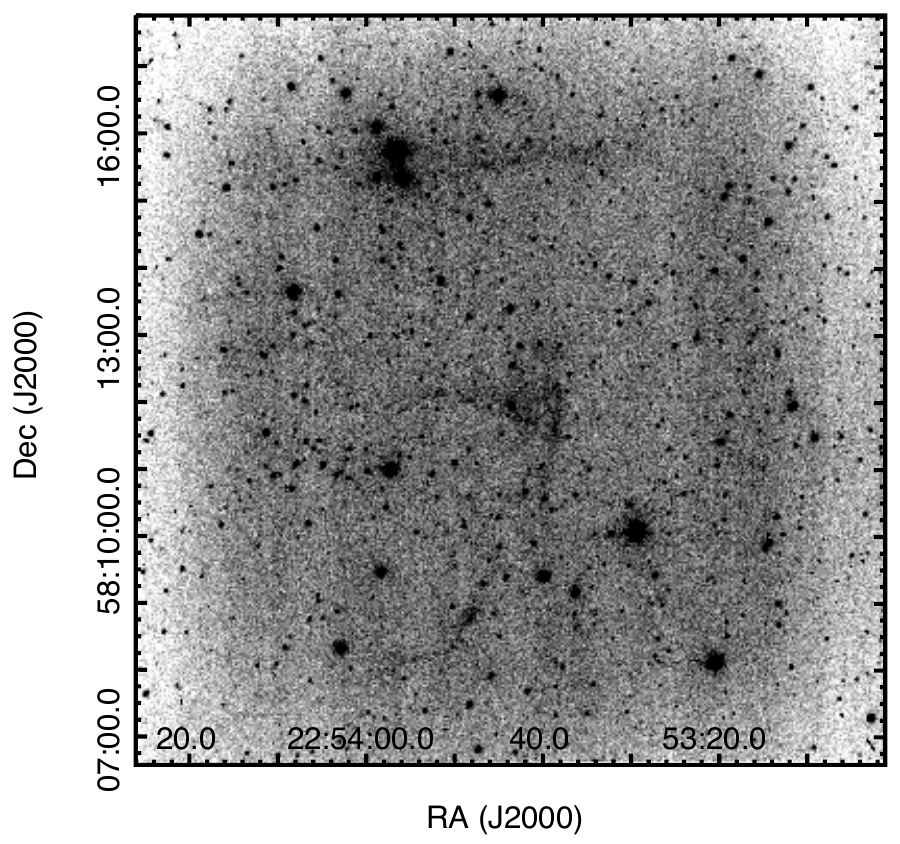}
\caption{The H$\alpha$ image of S region (S1: left and S2: right) of G108.2$-$0.6.}
\label{figure3}
\end{figure*}

We detected optical emission from N (three locations, namely N1, N2, and N3) and S (two locations, namely S1 and S2) regions of G108.2$-$0.6 using the H$\alpha$ filter.    
We show the position of these regions in Fig. \ref{figure1} with black boxes. The H$\alpha$ and continuum-subtracted H$\alpha$ images of the N region are given in Fig. \ref{figure2}. We show the H$\alpha$ images of the S region in Fig. \ref{figure3}. Due to bad weather conditions, we could not observe the S region using the H$\alpha$ continuum filter.

\subsubsection{Spectra}
\begin{figure*}
\includegraphics[angle=0, width=8.2cm]{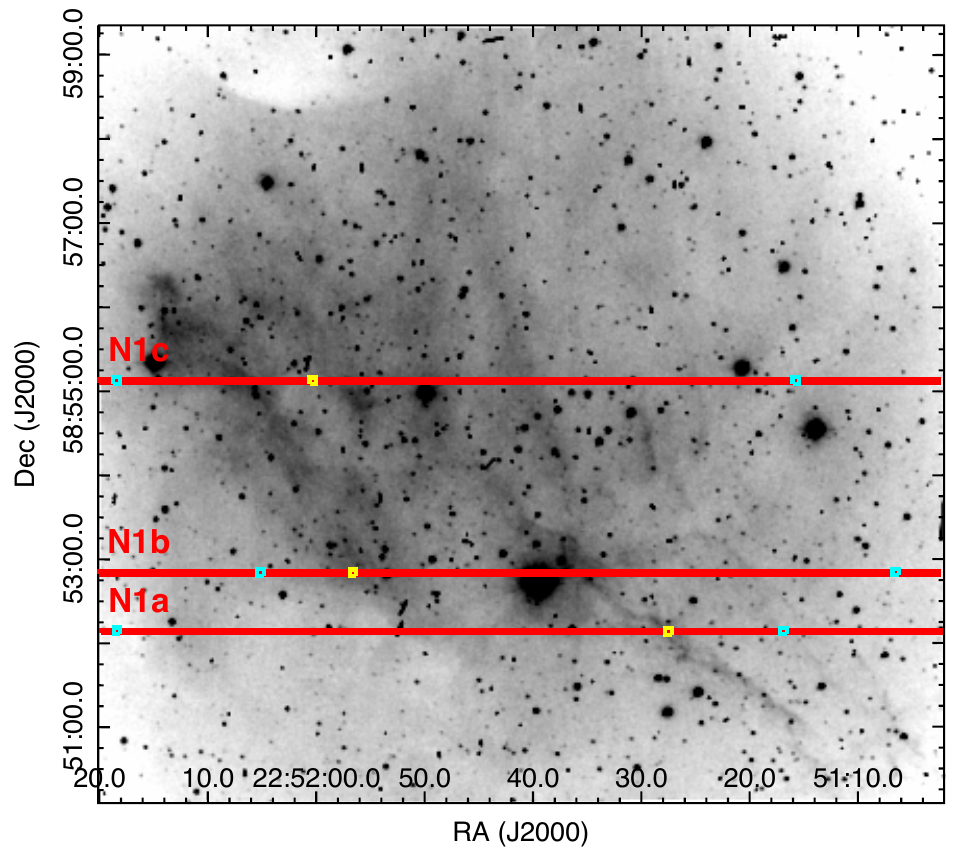}
\includegraphics[angle=0, width=8.0cm]{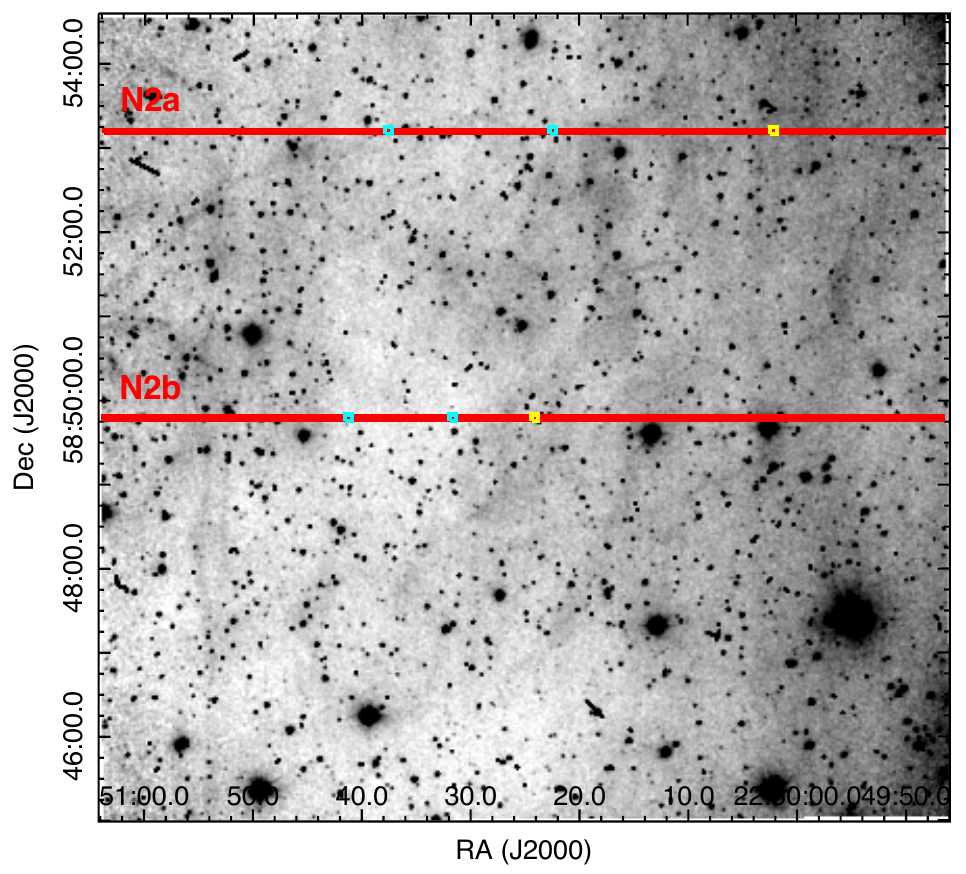}
\includegraphics[angle=0, width=8.0cm]{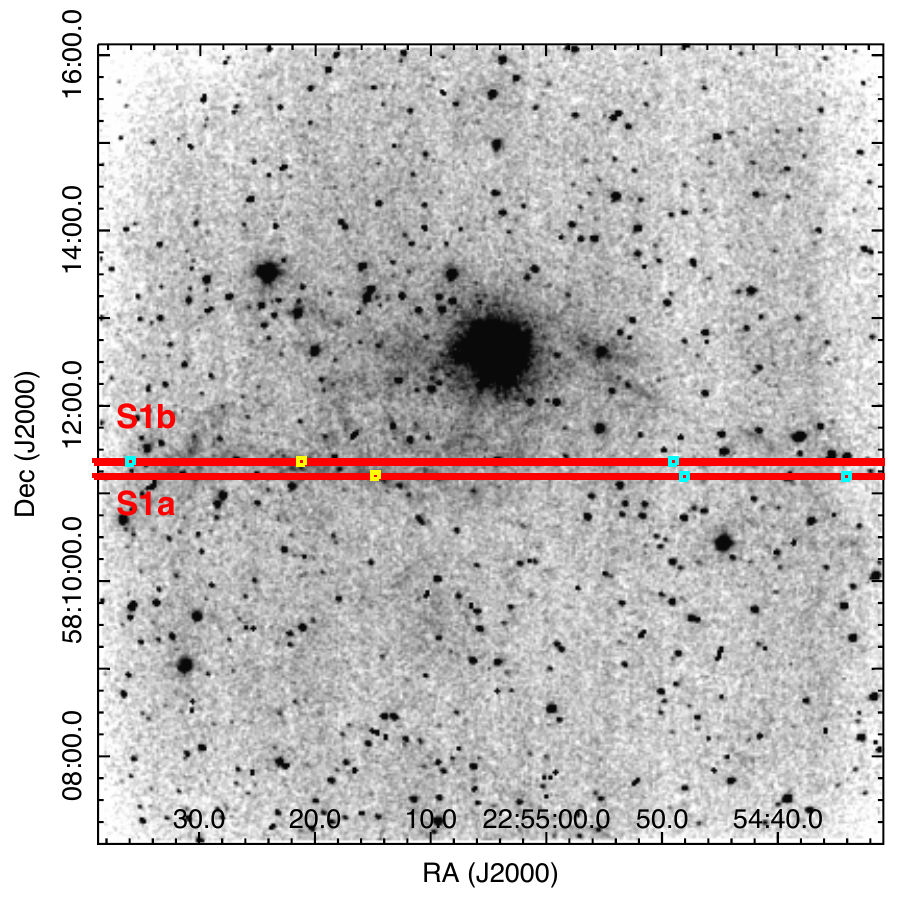}
\includegraphics[angle=0, width=8.2cm]{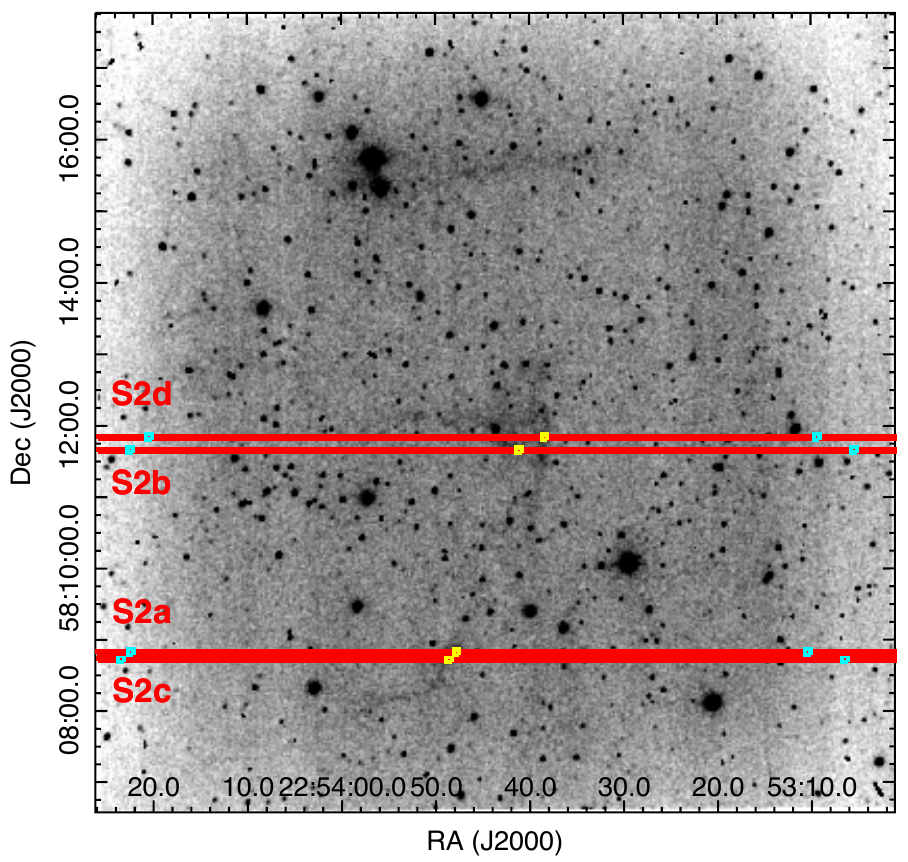}
\caption{The location of the slits is shown on the figures. Additionally, the regions where the SNR spectra were extracted are marked with yellow boxes, and the locations for the sky subtraction along each slit are marked with cyan boxes.}
\label{slits}
\end{figure*}

Our long-slit spectra were taken on the bright optical filaments and diffuse emission in the N and S regions of the SNR (their locations are given in Table \ref{Table2}). In Fig. \ref{slits}, we also show the slit locations on the H$\alpha$ images.

\begin{figure*}
\includegraphics[angle=0, width=8.5cm]{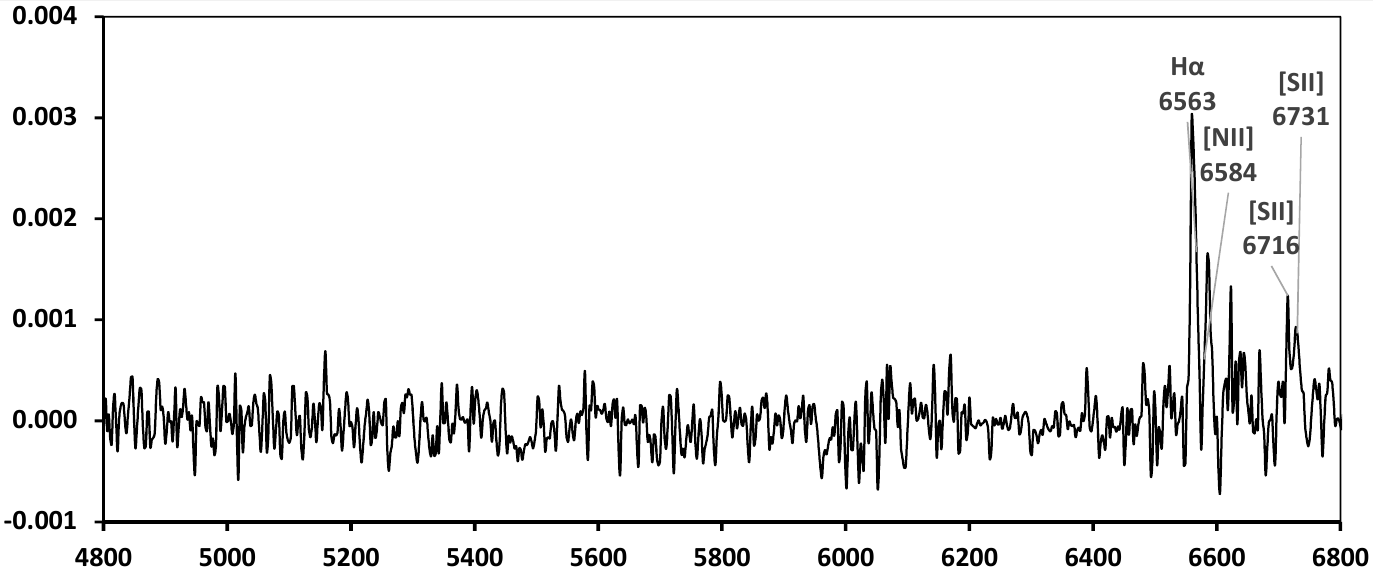}
\includegraphics[angle=0, width=8.5cm]{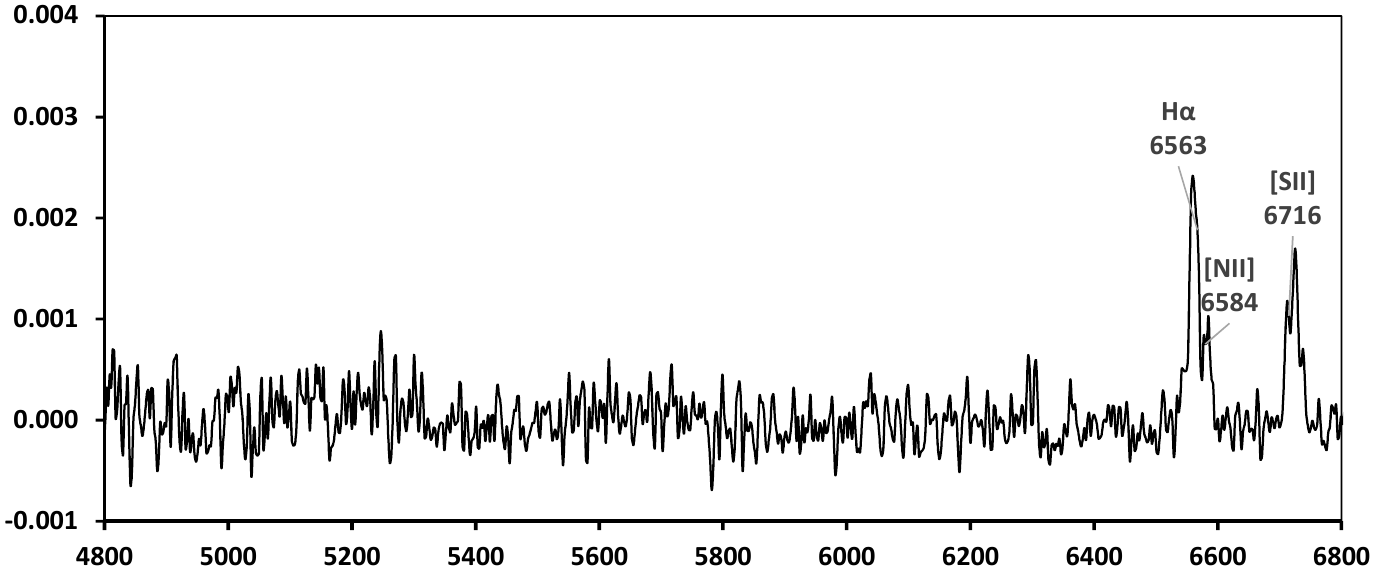}
\includegraphics[angle=0, width=8.5cm]{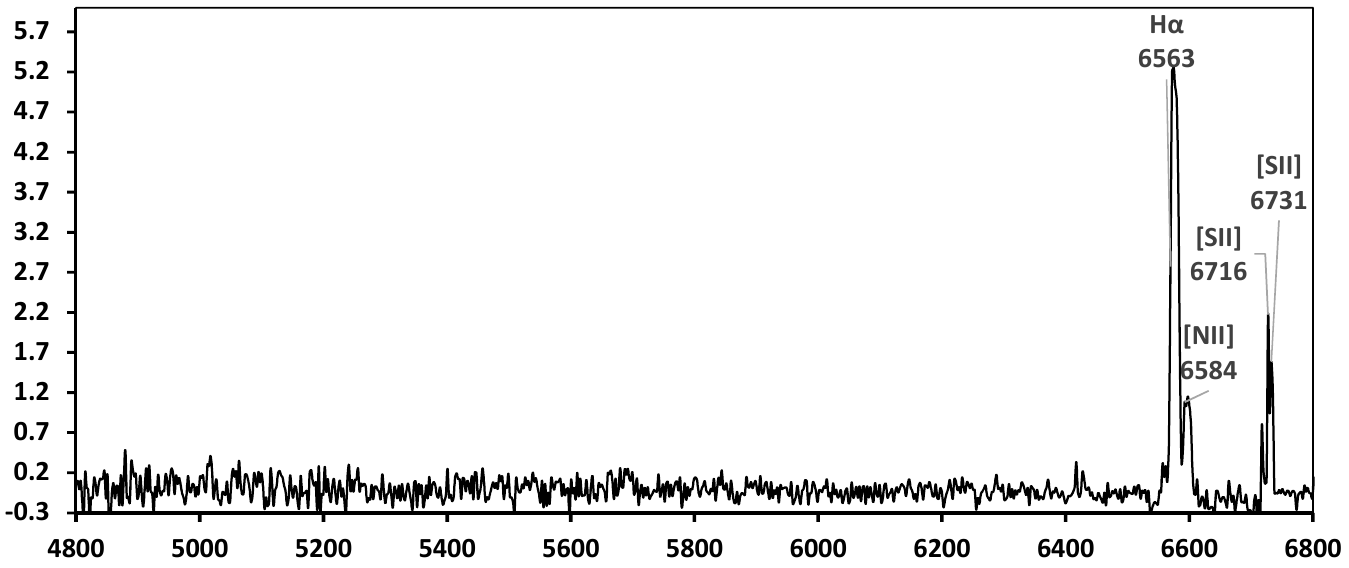}
\includegraphics[angle=0, width=8.5cm]{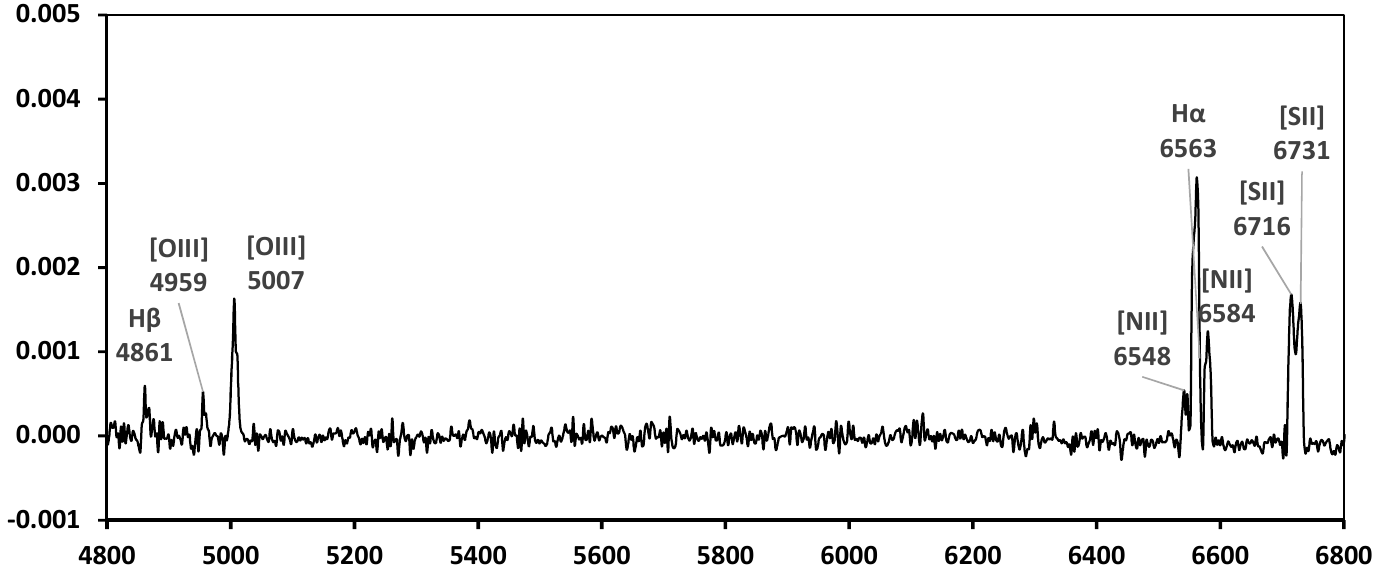}
\includegraphics[angle=0, width=8.5cm]{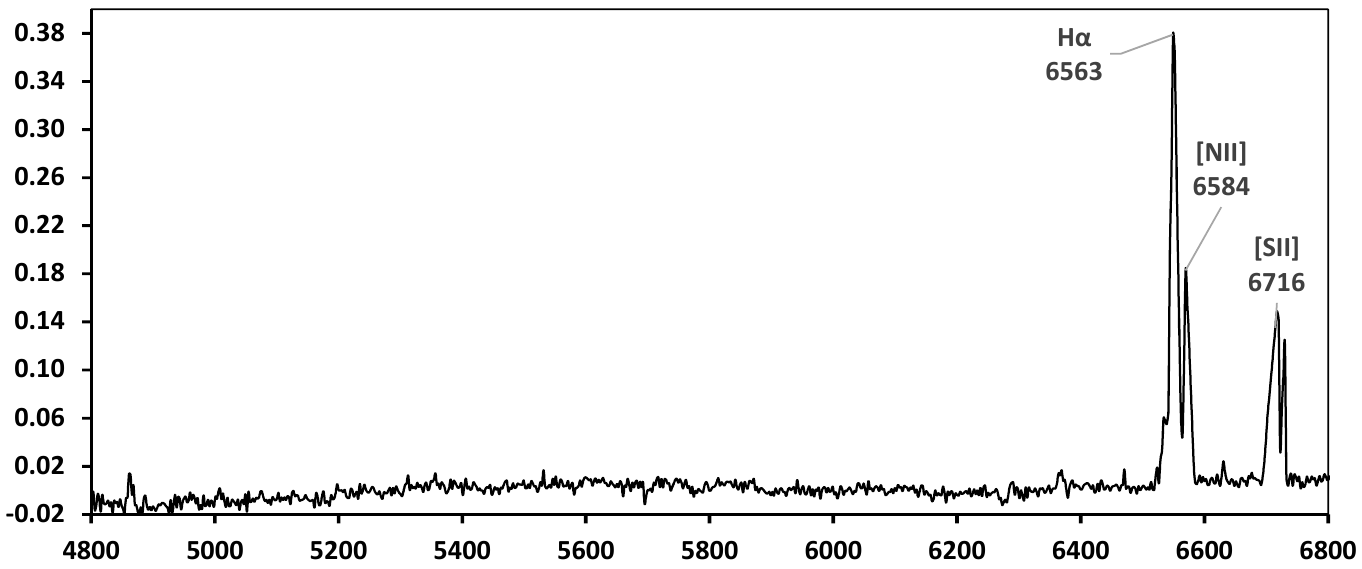}
\caption{Long-slit spectra of five locations (N1a, N1b, N1c, N2a, and N2c) in N region of G108.2$-$0.6 showing the H$\beta$$\lambda$4861, H$\alpha$$\lambda$6563, [O\,{\sc iii}]$\lambda$4959, $\lambda$5007, [N\,{\sc ii}]$\lambda$6548, $\lambda$6584 and [S\,{\sc ii}]$\lambda$6716, $\lambda$6731 lines. See Table \ref{Table2} for slit locations. The y-axis shows the flux in $10^{-14}$ ergs cm$^{-2}$ s$^{-1}$ \AA$^{-1}$, while the x-axis shows the wavelength in Angstroms.}
\label{figure4}
\end{figure*}

\begin{figure*}
\includegraphics[angle=0, width=8.5cm]{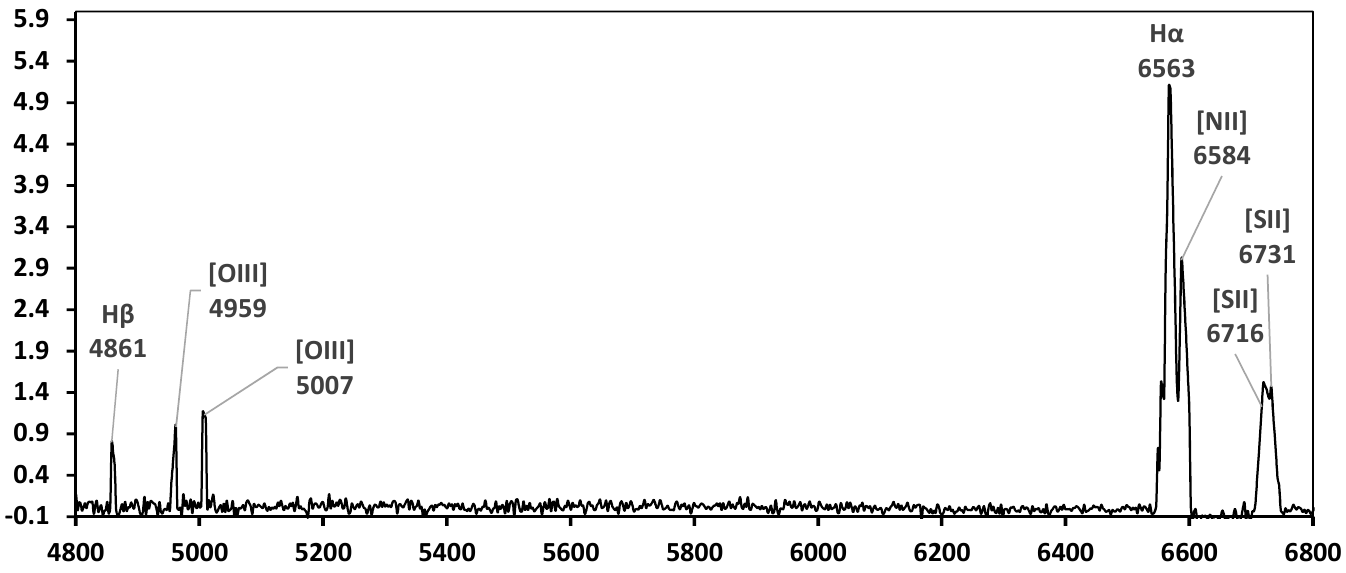}
\includegraphics[angle=0, width=8.5cm]{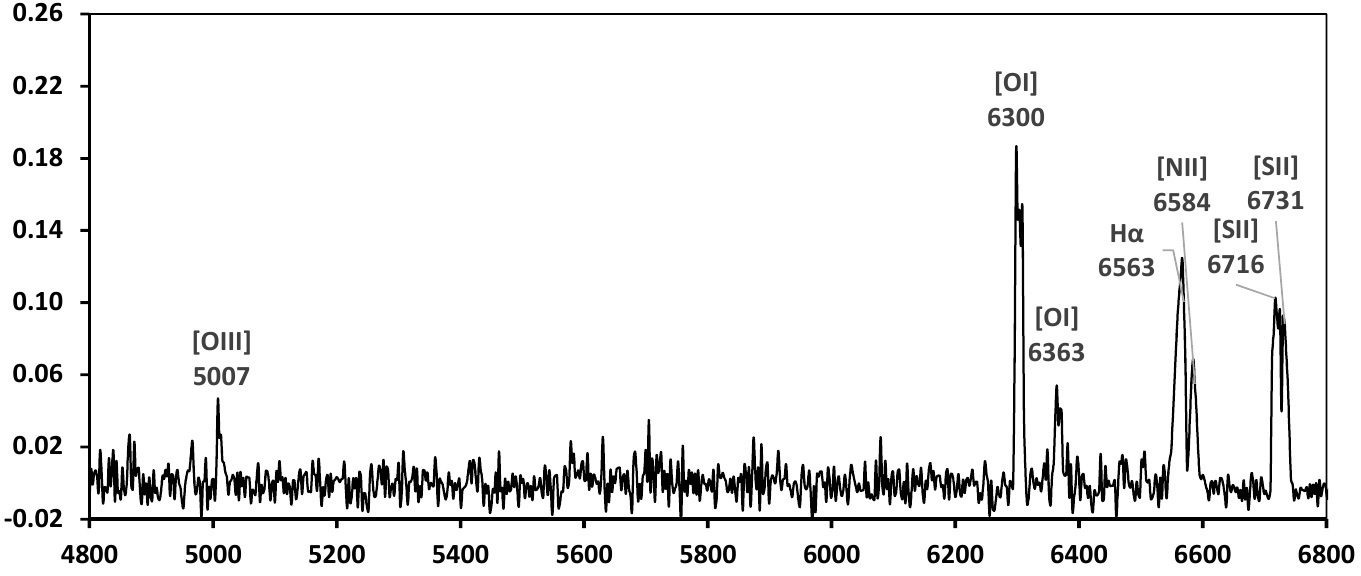}
\includegraphics[angle=0, width=8.5cm]{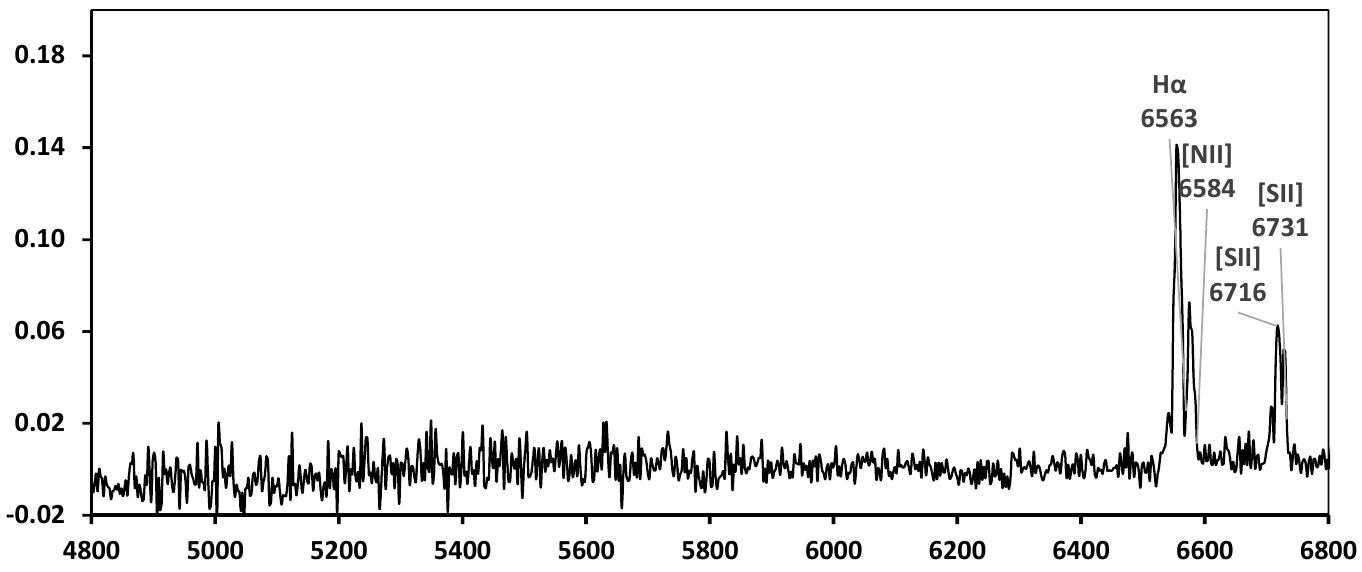}
\includegraphics[angle=0, width=8.5cm]{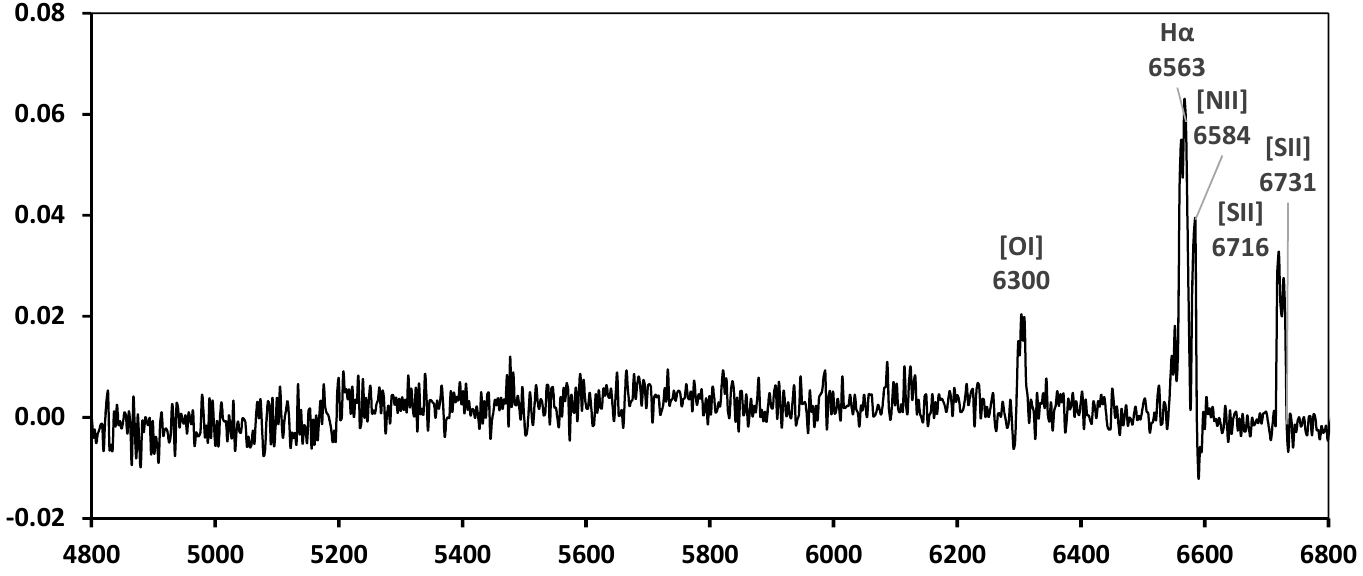}
\includegraphics[angle=0, width=8.5cm]{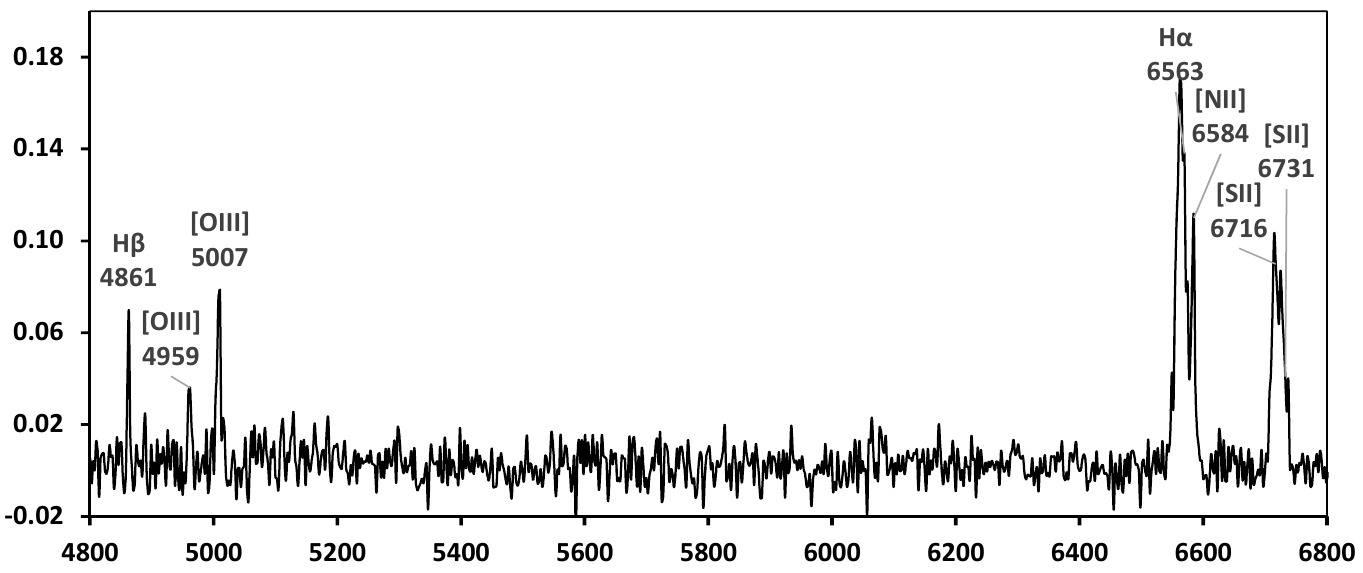}
\includegraphics[angle=0, width=8.5cm]{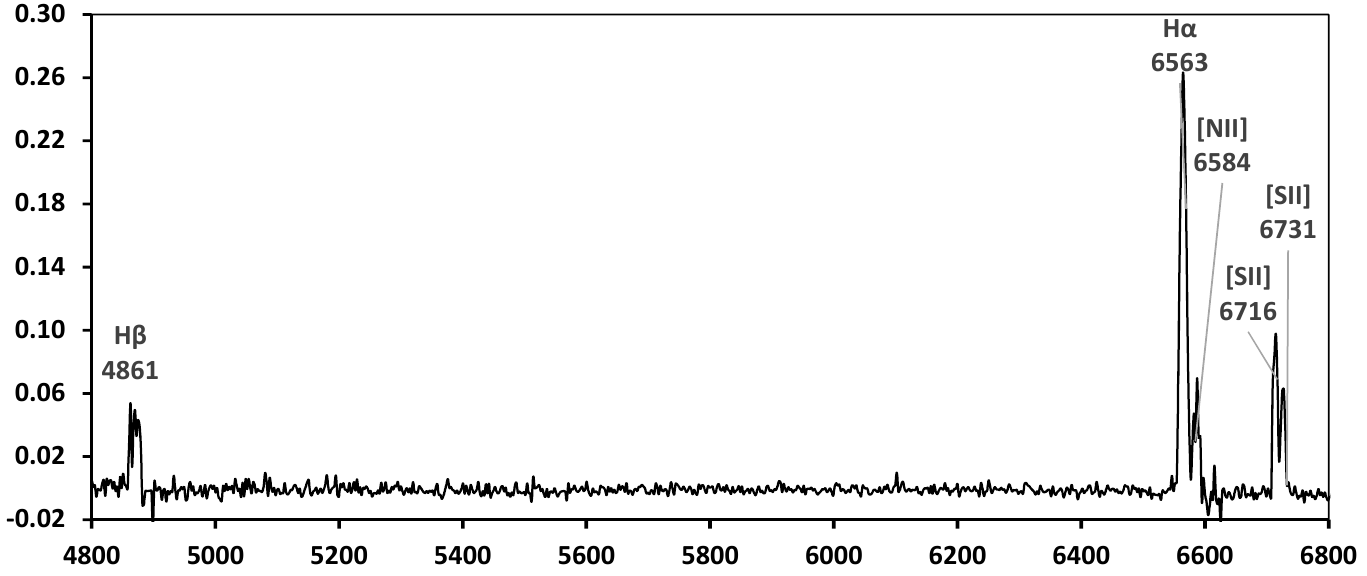}
\caption{Long-slit spectra of six locations (S1a, S1b, S2a, S2b, S2c, and S2d) in S region of G108.2$-$0.6 showing the H$\beta$$\lambda$4861, H$\alpha$$\lambda$6563, [O\,{\sc iii}]$\lambda$4959, $\lambda$5007, [O\,{\sc i}]$\lambda$6300, $\lambda$6363, [N\,{\sc ii}]$\lambda$6584 and [S\,{\sc ii}]$\lambda$6716, $\lambda$6731 lines. See Table \ref{Table2} for slit locations. The y-axis shows the flux in $10^{-14}$ ergs cm$^{-2}$ s$^{-1}$ \AA$^{-1}$, while the x-axis shows the wavelength in Angstroms.}
\label{figure5}
\end{figure*}

The long-slit spectra in the 4800$-$6800 {\AA} range are presented in Figs \ref{figure4} and \ref{figure5}, respectively. Line fluxes and their $1\sigma$ errors were determined using the deblending function in the \texttt{splot} task in \texttt{IRAF}. The \texttt{rms} was measured on either side of the emission feature to compute errors in the measured line fluxes and then averaged. This average rms was then set as $\sigma_0$ in the \texttt{splot} error package, with \texttt{nerrsample} set to 100 and \texttt{invgain = 0.} The line fluxes and line ratios are given in Table \ref{Table3}. 

 \begin{table*}
\centering
\caption{Line fluxes (normalised to $F$(H$\alpha$) = 100) with the $1\sigma$ errors. The line ratios are also presented.}
\label{Table3}
 \begin{tabular}{@{}p{3.5cm}p{1.8cm}p{1.8cm}p{1.8cm}p{1.8cm}p{1.8cm}p{1.8cm}@{}}
 \hline
 \hline
  &	 N1a	&	N1b	&	 N1c	&	 N2a  &	 N2b	 \\[0.5 ex]
\hline												
H$\beta$ ($\lambda$4861)	&	$-$	&	$-$	&	$-$	&	13 $\pm$ 2 &	$-$	  \\
												
$[$O$\,${\sc iii}$]$ ($\lambda$4959) 	&	$-$	&	$-$	&	$-$	&	11 $\pm$ 2	&	$-$	   \\
												
$[$O$\,${\sc iii}$]$ ($\lambda$5007) 	&	$-$	&	$-$	& $-$	&  42 $\pm$ 2 	&	$-$   \\
												
$[$O$\,${\sc i}$]$ ($\lambda$6300) 	&	$-$	&	$-$	&	$-$	&	$-$	&	$-$ \\
												
$[$O$\,${\sc i}$]$ ($\lambda$6363) 	&	$-$	&	$-$	&	$-$	&	$-$	&	$-$ \\
												
$[$N$\,${\sc ii}$]$ ($\lambda$6548) 	&	 $-$	&	$-$	&	$-$	& 15 $\pm$ 3	&	$-$  \\
												
H$\alpha (\lambda$ 6563) 	&	100 $\pm$ 6	&	100 $\pm$ 10	&	100 $\pm$ 2	&	100 $\pm$ 3 &	100 $\pm$ 2	  \\
												
$[$N$\,${\sc ii}$]$ ($\lambda$6584) 	&	53 $\pm$ 8	&	15 $\pm$ 5 	&	19 $\pm$ 2	&	34 $\pm$ 3 &	26 $\pm$ 2   \\
												
$[$S$\,${\sc ii}$]$ ($\lambda$6716) 	&	25 $\pm$ 9	&	62 $\pm$ 5 	&	$-$ &	48 $\pm$ 3 &	48 $\pm$ 1   \\
												
$[$S$\,${\sc ii}$]$ ($\lambda$6731) 	&	33 $\pm$ 13	&	$-$	&
25 $\pm$ 2	&	45 $\pm$ 4 &	$-$ \\[0.5 ex]

F (H$\alpha$) (erg cm$^{-2}$ s$^{-1}$) &  3.44$\times$$10^{-16}$  & 6.20$\times$$10^{-16}$ & 8.00$\times$$10^{-13}$ & 5.66$\times$$10^{-17}$ & 6.22$\times$$10^{-14}$  \\
												
[S\, {\sc ii}]/ H$\alpha$ 	&	   0.58 $\pm$ 0.19 &	   0.62 $\pm$ 0.02	&	   0.25 $\pm$ 0.02	&	   0.93 $\pm$ 0.04	& 0.48 $\pm$ 0.01  \\[0.5 ex]

$n_{\rm e}$(cm$^{-3}$)	&	    1800 $\pm$ 150	&	  $-$	&	     $-$ 	&	 530 $\pm$ 40 	&  $-$ \\
	\\																					
  \hline												
 \hline												
        &	 S1a 	&	 S1b	&	 S2a	&	 S2b &	 S2c	&	 S2d	 \\[0.5 ex]
\hline												
												
H$\beta$ ($\lambda$4861)	&	5 $\pm$ 2	&	$-$	&	$-$	&	$-$ &	9 $\pm$ 2	&	26 $\pm$ 1   \\
												
$[$O$\,${\sc iii}$]$ ($\lambda$4959) 	&	6 $\pm$ 3	&	$-$	&	$-$	&	$-$ &	8 $\pm$ 2	&	$-$  \\
												
$[$O$\,${\sc iii}$]$ ($\lambda$5007) 	&	8 $\pm$ 3	&	15 $\pm$ 8	&	$-$	&	$-$ &	20 $\pm$ 2 	&	$-$ \\
												
$[$O$\,${\sc i}$]$ ($\lambda$6300) 	&	$-$	&	104 $\pm$ 3 	&	$-$	&	23 $\pm$ 2 &	$-$	&	$-$	 \\
												
$[$O$\,${\sc i}$]$ ($\lambda$6363) 	&	$-$	&	26 $\pm$ 3 	&	$-$	&	$-$	&	$-$	&	$-$	   \\
												
$[$N$\,${\sc ii}$]$ ($\lambda$6548) 	&	$-$	&	$-$	&	$-$	&	$-$	&	$-$	&	$-$	  \\
												
H$\alpha (\lambda$ 6563) 	            &	100 $\pm$ 2	&	100 $\pm$ 8	&	100 $\pm$ 7	&	100 $\pm$ 8 	&	100 $\pm$ 6	&	100 $\pm$ 2  \\
												
$[$N$\,${\sc ii}$]$ ($\lambda$6584) 	&	33 $\pm$ 5 &	30 $\pm$ 5	&	30 $\pm$ 4	&	24 $\pm$ 4	&	24 $\pm$ 3	&	21 $\pm$ 2	   \\
												
$[$S$\,${\sc ii}$]$ ($\lambda$6716) 	&	23 $\pm$ 2	&	62 $\pm$ 5 &	24 $\pm$ 3	&	23 $\pm$ 6 &	28 $\pm$ 5	&	25 $\pm$ 1	 \\
												
$[$S$\,${\sc ii}$]$ ($\lambda$6731) 	&	23 $\pm$ 2	&	50 $\pm$ 6 &	17 $\pm$ 2	&	19 $\pm$ 5 &	36 $\pm$ 6	&	17 $\pm$ 1	  \\[0.5 ex]

F (H$\alpha$) (erg cm$^{-2}$ s$^{-1}$) &  1.14$\times$$10^{-12}$ & 2.32$\times$$10^{-14}$ & 2.25$\times$$10^{-14}$ &1.05$\times$$10^{-14}$  &	3.37$\times$$10^{-14}$	&	3.38$\times$$10^{-14}$	\\
												
[S\, {\sc ii}]/ H$\alpha$ 	&	    0.46 $\pm$ 0.03  	&	  1.12 $\pm$ 0.02  	&	   0.41 $\pm$ 0.02   	&	   0.42 $\pm$ 0.08   &	0.64 $\pm$ 0.07 	&	0.42 $\pm$ 0.01 	   \\[0.5 ex]
												
$n_{\rm e}$(cm$^{-3}$)	&	    680 $\pm$ 20	&	 240 $\pm$ 40	&	   60 $\pm$ 20 	&	 280 $\pm$ 60	&	 1640 $\pm$ 90  &	 15 $\pm$ 10	  \\
\\[0.5 ex]											
  \hline
 \hline
     
\end{tabular}
\end{table*}

\subsection{H\,{\sc i} analysis and results}
\begin{figure*}
\includegraphics[angle=0, width=80mm]{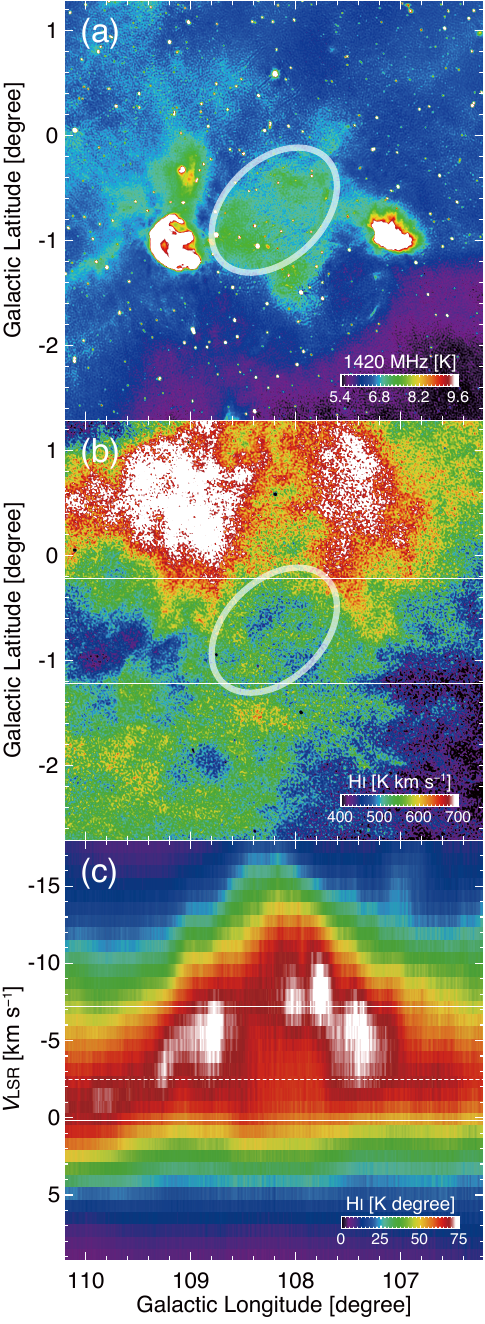}
\caption{(a) Large-scale map of radio continuum at 1420 MHz toward G108.2$-$0.6 \citep{Ta03}. (b) Velocity-integrated intensity map of H{\sc i} at $V_\mathrm{LSR} = -8.9$--0.2~km~s$^{-1}$. The superposed circles indicate the shell boundary of the SNR. (c) Position--velocity diagram of H{\sc i}. The integration range of Galactic latitude is from $-1.21^{\circ}$ to $-0.21^{\circ}$.}
\label{figure7}
\end{figure*}

According to \citet{Ti07}, the H\,{\sc i} features in the velocity range from -58 to -53~km~s$^{-1}$ show suggestive correlations with the radio continuum shell of SNR G108.2$-$0.6. On the other hand, our analysis of the same H\,{\sc i} data found another counterpart of H\,{\sc i} which is possibly associated with the SNR. Fig. \ref{figure7}(b) shows the H\,{\sc i} integrated intensity map at the velocity range from -8.9 to 0.2~km~s$^{-1}$, superposed on the radio shell boundary as defined by eye inspection (Fig. \ref{figure7}(a)). We found a cavity-like distribution of H\,{\sc i} which shows a nice spatial correspondence with the radio shell boundary. Moreover, we newly found possible evidence for the expanding shell of H\,{\sc i} at the same velocity range. Fig. \ref{figure7}(c) shows the position-velocity diagram of H\,{\sc i}. We can see a hollowed-out distribution of H\,{\sc i}, whose spatial extent of galactic longitude is roughly consistent with that of the SNR. The H\,{\sc i} clouds outside of the SNR shell show a constant central velocity at $\sim$3.5 km s$^{-1}$.

\clearpage

\section{Discussion}
\label{discuss}
\subsection{Optical morphology}
Our H$\alpha$ survey showed the existence of filamentary and diffuse structures, which are clearly seen and well correlate  with the radio morphology as shown in Figs \ref{figure1}$-$\ref{figure3}.  The spatial correspondence of the optical emission with the radio is the strongest morphological evidence for the emission being due to the SNR. 

The optical emission from G108.2$-$0.6 is mostly diffuse (see Figs \ref{figure1}$-$\ref{figure3}). The detection of several filaments is consistent with the SNR emission. As can be seen in the top panel of Fig.~\ref{figure2},  several bright, long and parallel H$\alpha$ filaments located in the N1 region of the SNR. The bright diffuse emission is seen in the N2 and N3 regions (see medium and bottom panels of Fig.~\ref{figure2}). Much shorter and curved filaments are visible in the S1 and S2 regions (see Fig.~\ref{figure3}). We note that our S1 region is close to the northeast region of SNR G107.5$-$1.5 (see \citealt{Ko03, Ba23}). 

G108.2$-$0.6 is a large SNR, and its long and curved filament structure supports the shocks, which are expanding into a large-scale local ambient medium with varying pre-shock densities.

\subsection{Optical spectral properties}
The spectra exhibit [S\,{\sc ii}]/H$\alpha$ ratios in range 0.4$-$1.1, indicating emission from shock-heated gas (e.g. \citealt{Fe85}), except N1c spectrum ([S\,{\sc ii}]/H$\alpha$ $\sim$ 0.25) which is consistent with the photoionized gas \citep{MaFe97, BlLo97}. 

Using the \texttt{temden} routine and assuming a temperature of 10$^{4}$ K, we calculated the electron density is ranging from 15 to 1800 cm$^{-3}$) from the [S\,{\sc ii}] 6716/6731 ratio \citep{OsFe06}. The high $n_{\rm e}$ values  ($\sim$1800 and $\sim$1640 cm$^{-3}$) are found for the N1a and S2c slit locations can be attributed to the interaction of the SNR with a dense medium.

For the N2a, S1a, and S2c slit locations, we used the $[$O$\,${\sc iii}$]$ $\lambda$5007/H$\beta$ ratio and a planar shock model of \citet{Ha87} to estimate the shock velocity $V_{\rm s}$  of  $\sim$80 km s$^{-1}$. Other slit locations lack $[$O$\,${\sc iii}$]$ emission, indicating shock velocities less than about $\sim$70 km s$^{-1}$.
 
We then derived the pre-shock cloud density ($n_{\rm c}$) to be $\sim$18$-$57 cm$^{-3}$ taking the relation 
\begin{equation}
n_{\rm [S\,{\sc II}]} = 45~n_{\rm c} \times (V/100~{\rm km~s^{-1}})^{2}~~~{\rm cm}^{-3}
\end{equation}
from \citet{Fe80}, where $n_{\rm [S\,{\sc II}]}$ is the electron density calculated from the sulfur line ratio.

We detected H$\beta$ emission in the N2a, S1a, S2c, and S2d locations and therefore can estimate the extinction in these directions. We found extinction $E(B-V)$ value of $\sim$0.22$-$1.65 using the H$\alpha$/H$\beta$ ratio  and assuming that $E(B-V)=0.664c$, where $c$ is a logarithmic extinction (see \citealt{Ka76, Al84}).  We finally estimated the column density $N_{\rm H}$ of  $\sim$ (1.2$-$8.9)$\times$ $10^{21}$ cm$^{-2}$ using the relation $N_{\rm H}$ = 5.4 $\times$ $10^{21}$ $\times$ $E(B-V)$ \citep{Pr95}. The extinction and column density values show significant variation across the observed regions.

Overall, our optical spectral investigation shows that G108.2$-$0.6 is expanding into dense and complex areas, which agrees well with the conclusion reported in \citet{Ti07}. 

\subsection{Atomic environment}
Our analysis found alternative counterparts of H\,{\sc i} clouds ($V_\mathrm{LSR}$: -8.9$-$0.2 ~km~s$^{-1}$) possibly associated with the SNR. The cavity-like structures of H\,{\sc i} shown in Figs \ref{figure7}(b) and \ref{figure7}(c) are likely an expanding H\,{\sc i} shell which was generally formed by strong stellar winds from the progenitor and/or supernova shocks (e.g., \citealt{Ko90, Ko91}). When we adopt the systemic velocity of the H\,{\sc i} clouds to be $\sim$3.5~km~s$^{-1}$, the expanding velocity of the H\,{\sc i} cloud gives $\sim$5~km~s$^{-1}$, which is roughly consistent with other Galactic/Magellanic SNRs (e.g., \citealt{Sano17, Sano19, Ku18}). The shock front of the supernova is now impacting on the inner wall of the H\,{\sc i} expansion shell.

\subsection{Distance and Age}
If the H\,{\sc i} cloud at $V_\mathrm{LSR}$ = -8.9$-$0.2 km s$^{-1}$ is physically associated with SNR G108.2$-$0.6, we may reconsider the distance to the SNR. By adapting the Galactic rotation model with the IAU-recommended values ($R_{0}$ = 8.5 kpc, $\Theta_{0}$ = 220~km~s$^{-1}$; 
\citealt{Ke86, Br93}), we obtain a kinematic distance of the H\,{\sc i} cloud and SNR G108.2$-$0.6 as $\sim$0.8 kpc. This value is roughly consistent with the extinction distance \citep{Zh20}. At the distance, we derived the diameter of SNR G108.2$-$0.6 to be 16.3 pc $\times$ 12.6 pc. By assuming the Sedov–Taylor model \citep{Se59}, the dynamical age of SNR G108.2$-$0.6 can be estimated to be $\sim$70$\pm$10 kyr.

\section{Conclusions}
\label{conc}
In this work, we report the first detection of optical emission from G108.2$-$0.6 based on both H$\alpha$ imaging and spectroscopic observations. The filamentary structure of G108.2$-$0.6 is consistent with its SNR nature. Our spectra exhibit the H$\beta$$\lambda$4861, [O\,{\sc iii}]$\lambda$4959, $\lambda$5007, [O\,{\sc i}]$\lambda$6300, $\lambda$6363, H$\alpha$$\lambda$6563, [N\,{\sc ii}]$\lambda$6584 and [S\,{\sc ii}]$\lambda$6716, $\lambda$6731 lines. The [O\,{\sc i}]$\lambda$6300, $\lambda$6363 lines detected in the S region also support the indicator of the presence of shocks. The [S\,{\sc ii}]/H$\alpha$ ratios suggest that the shock-excited gas for G108.2$-$0.6. We found the electron density ranging from 15 to 1800 cm$^{-3}$ using [S\,{\sc ii}] 6716/6731. The spectra show a relatively low shock velocity of $V_{\rm s}$ $\sim$ 80 km s$^{-1}$ with the pre-shock cloud density of $n_{\rm c}$ $\sim$18$-$57 cm$^{-3}$. We also estimated the kinematic distance to the SNR of $\sim$0.8 kpc and the dynamical age as $\sim$70$\pm$10 kyr using the archival H\,{\sc i} data. Further observations with different frequencies are needed to better understand the nature of this SNR. 

\section*{Acknowledgements}
We thank T\"{U}B\.{I}TAK National Observatory for partial support in using RTT150 with project number 1562. HS was supported by JSPS KAKENHI grant No. 21H01136. The Canadian Galactic Plane Survey (CGPS) is a Canadian project with international partners. The Dominion Radio Astrophysical Observatory is operated as a national facility by the National Research Council of Canada. The CGPS is supported by a grant from the Natural Sciences and Engineering Research Council of Canada. We also thank the referee for valuable comments and suggestions that helped to improve the paper.

\section*{DATA AVAILABILITY}
The optical data from the RTT150 telescope will be shared on reasonable request to the corresponding author.

 


\bsp	
\label{lastpage}
\end{document}